# Drag and Drop: Influences on the Design of Reusable Software Components

B.Jalender
Asst.Professor
IT Dept, VNRVJIET
Hyderabad, INDIA.

Dr A.Govardhan
Principal, JNTUCEJ
Jagtial, Karimnagar
A.P, INDIA.

Dr P.Premchand
Professor, CSE Dept
Osmania University
Hyderabad, INDIA.

Dr C.Kiranmai
Professor
CSE Dept, VNRVJIET
Hyderabad, INDIA

G.Suresh Reddy
Associate Professor
IT Dept, VNRVJIET
Hyderabad, INDIA.

Department of Information Technology
VNRVJIET, Hyderabad, Andhra Pradesh, INDIA – 500090.

*Abstract*— The fundamental unit of large scale software construction is the component. A component is the fundamental user interface object in Java. Everything you see on the display in a java application is a component. The ability to let users drag a component from the Interface and drop into your application is almost a requirement of a modern, commercial user interface. The CBD approach brings high component reusability and easy maintainability, and reduces time-to-market. This paper describes the component repository which provides functionality for component reuse process through the drag and drop mechanism and it's influences on the reusable components..
*(Abstract)*

## I. INTRODUCTION

Component is fundamental unit of large scale software construction. Every component has an interface and an Implementation [1].The interface of a component is *anything* that is visible externally to the component. Everything else belongs to its implementation. A component is the fundamental user interface object in Java. Everything we seen on the display in a Java application is a component. This includes things like windows, drawing canvases, buttons, checkboxes, scrollbars, lists, menus, and text fields [2]. To be used, a component usually must be placed in a container. Container objects group components, arrange them for display using a layout manager, and associate them with a particular display device. The Component Based Development approach brings high component reusability and easy maintainability, and reduces time-to-market [6]. Therefore it improves productivity of software systems and lower development cost [1],[2].In the context of reusable software components, copying leads to two classes of difficulties: components whose implementations are inherently inefficient, and client programs that are hard to reason about [4].

As a component is unit software that has business logics and interfaces, the component communicate with other omponents through its interfaces. Component-Based Development (CBD) approach develops software systems by assembling preexisting components under well-defined architecture or framework [1],[2]. The CBD approach brings high component reusability and easy maintainability, and reduces time-to-market. Therefore it improves productivity of software systems and lower development cost [2],[3]. To develop component-based software, software developers face a challenge to find software components among the components that are previously built. When the number of component grows, the complexity of components becomes greater. Hence, management for the existing components is basically required.

To implement a range of services in component based software, firstly a set of compatible components are identified from existing components [3]. The component that doesn't operate correctly together should be modified according to dependencies between other components. Components with different form from a component are generated, which need to test modified functionality to check correctness of the changes. In developing component-based software, the activities such as component identification, component modification, and component test make consequence for component reuse.

Until the Java 2 platform hit the streets, drag and drop support (specifically support for interacting with the native windowing system underneath the JVM[7]) was lacking. The ability to let users drag a file from their file choosers into your application is almost a requirement of a modern, commercial user interface. The java.awt.dnd package gives you and your Java programs access to that support. Now you can create applications that accept information dropped in from an outside source. You can create Java programs that build up draggable information that you export to other applications. And of course, you can add both the drop and the drag capabilities to a single application to make its interface much more rich and intuitive.

In this paper, we provide a GUI interface that makes convenient access to the components by the Drag and Drop mechanism, and reuse those components after dragging.

This paper is organized as follows. Firstly, we describe what is Drag and Drop in Section 1 and Section 2 explains Proposed work section 3 mentions results of the Drag and Drop mechanism. Finally, Section 4 describes conclusion and future work.

If you have ever used a graphical file system manager to move a file from one folder to another, you have used drag and drop (often abbreviated DnD). The term "drag and drop" refers to a GUI action whereby the end user "picks up" an object such as a file or piece of text by pressing the mouse button down on that object[4]. Then without letting up on the button, the user "drags" it over some area of the screen and lets up on the mouse button to "drop" the object.

In this paper we work with two main areas a drop target, and a drag source. A drop target accepts an incoming drag. The process of accepting the dragged information generates a series of events that we can respond to. Figure(6) represent the GUI with drop target. The source of the dragged item might be your application, another Java application, or some native window system application like your file manager or any component. The source doesn't matter to the drop target.

## Issues for drag and drop components

When you create a D&D-enabled component, there are several issues you need to address:

- **Starting the Drag operation**
- **Drag-under feedback**
- **Drag-over feedback**
- **The Drop**
- **Transferable**
- **The Move operation**

we will address each of these issues for the JTree[8] with the help of our D&D library classes[5]. First however, let's examine these issues in a general sense.

### Starting the Drag operation
First we have to check is it OK to start dragging at the current pointer location. Perhaps the component is a JTree and the selected tree node cannot be dragged. The DragGestureAdapter [5] object calls the isStartDragOk () method of our DragComponent object. If the return value is true, the drag simply will start.

### Drag-under feedback
To show that a drop is valid we defined the dragUnderFeedback()[5] and undoDragUnderFeedback()[5] methods from our DropComponent interface. This will make our GUI interface components are associated with the logical cursor.

### Drag-over feedback
Most of the time, the default drag-over feedback is fine. However, there could be a situation in which you would like to use custom cursors. The DropTargetAdapter[9] calls the DropComponent's isDragOk() method repeatedly during the drag. The return value sets the drag-over feedback. For most components the initial cursor should be a no-drop cursor, since dropping the data in the same place as its origin is useless[7].

### The Drop
The dropped data, is inserted it someplace in the component. In this aspect we provided a feature that the users drop components any where in our Interface [8].

### Transferable
The Transferable object may encapsulate data associated with the component or data retrieved from its model. Our DragGestureAdapter [9] calls the DragComponent's getTransferable() method.

### The Move operation
Because it's a two-step process, the move operation actually removes the data. The D&D system adds the components to the destination and then removes it from the source. If it's a complex component, such as a JList[9], The DragSourceAdapter[9] calls the DragComponent's move() method.

### Build a D&D library

If we need a single D&D-enabled component, we can create a subclass that defines DragGestureListener, DragSourceListener, and DropTargetListener as inner classes. If user needs a number of D&D-enabled components, user will write very similar code for each component's listeners.

A D&D-enabled component would create associations with an instance of each of these classes, data transfer in theDnD interfaceas shown in Figure.(1).

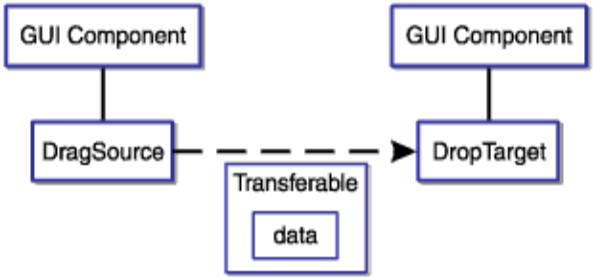

Figure 1. Data Transfer in DnD

**Drag classes**

A drag-enabled component implements the DragComponent interface. It creates an instance of DragGestureAdapter and an instance of DragSourceAdapter. The DragSourceAdapter implements the DragSourceListener interface and maintains a reference to a DragComponent object. When a drag is initiated, the DragSourceAdapter queries the DragComponent for the acceptable drag operation and an appropriate Transferable object(reusable component). If this is a move operation, the DragSourceAdapter will tell the DragComponent to move the component. The move operation actually adds the data to the destination, then removes the data from the source at the end of the D&D operation. These are usually cursor changes.

The DragComponent uses a DragGestureAdapter object, which implements the DragGestureListener interface, in order to initiate the component drag operation. With components such as a JTree, it is possible that not all nodes can be dragged. The DragGestureAdapter verifies the drag with the DragComponent's isStartDragOk() method, and it registers the DragComponent's DragSourceAdapter.

**The Drop classes**

A drop-enabled component implements the Drop Component interface. The implementation is shown in Below figure (2).

**Drag and Drop and Data Transfer**

Most programs can benefit from the ability to transfer information, either between components, between Java applications, or between Java and native applications Drag and drop (DnD) support. The following diagram illustrates the Java portion of a drag operation in our interface.

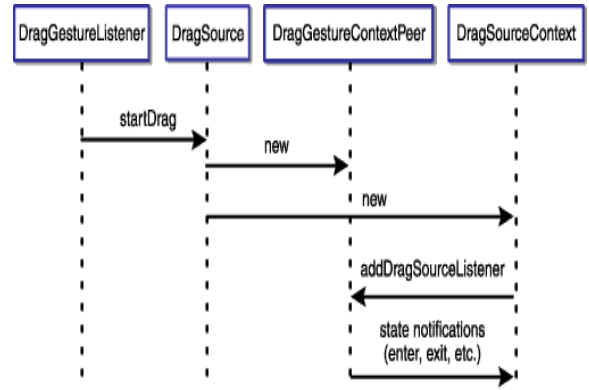

Figure 2. Drop Component Interfaces

.
**II. Proposed Work**

The Drag and drop functionality in our interface uses the same technique as most other GUI features drag source events, and drop events. To handle these events, we implement the corresponding listener interfaces. This process should sound familiar to anyone who has set up event handlers for other GUI components. For example, to respond to a dropped object (reusable component), we create an event handler that implements the DropTargetListener interface.

A DragSource comes into existence, associated with some presentation Component in the GUI, to initiate a Drag and Drop of some potentially Transferable of reusable components. The DragSource object manifests "Drag Over" feedback to the user, in the typical case by animating the GUI Cursor associated with the logical cursor.

**Data Flavors and Actions**

When the Transferable object (reusable component) encapsulates data, it makes the data available to DropTarget in a variety of DataFlavors. For a local transfer within the same JVM (Java virtual machine), Transferable provides an object reference. The results shown below are related to local transfer of reusable components with in the same JVM.

However, if we want transfer one reusable component to another JVM or to the native system, this wouldn't make any sense, so a DataFlavor using a java.io.InputStream subclass usually is provided. In this paper we transferred reusable components with in the same JVM.

When invoking a drag and drop operation of reusable components, you may request various drag and drop actions. The DnDConstants class defines the class variables for the supported actions:

**ACTION_NONE** → no action taken

**ACTION_COPY** → the DragSource leaves the data intact

**ACTION_MOVE** → the DragSource deletes the data upon successful completion of the drop

**ACTION_COPY** or **ACTION_MOVE** → the DragSource will perform either action requested by the DropTarget

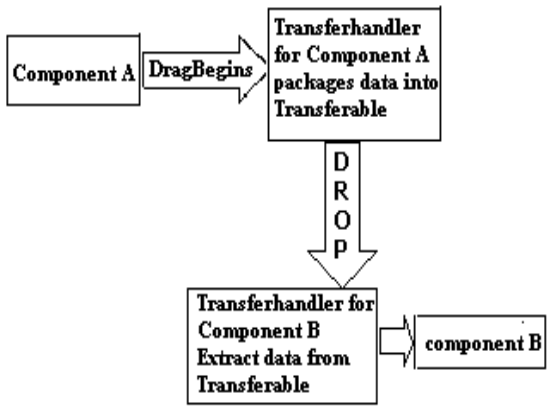

Figure 3.Drag Operation

The DragGestureListener causes the DragSource to initiate the Drag and Drop of components.The components from the GUI   interface are dropped into the Logical cursor.

**Cursor Icons for Drag and Drop**

| Microsoft Windows | Description |
| --- | --- |
|  | Copy. The destination underneath accepts reusable components. |
|  | Copy. The destination underneath will not accept reusable components. |
|  | Move. The destination underneath accepts reusable components. |
|  | Move. The destination underneath will not accept reusable componets. |

// **Implementation of DragGestureListener interface**.

```
 publicvoid dragGestureRecognized (DragGestureEvent dge) {
  // Get the mouse location and convert   it to a location within the tree.
   Point location = dge.getDragOrigin();
   TreePath dragPath = tree.getPathForLocation(location.x, location.y);
   if (dragPath != null && tree.isPathSelected(dragPath)) {
    // Get the list of selected files and create a Transferable
    // The list of files and the is saved for use when the drop completes.
    paths = tree.getSelectionPaths();
    if (paths != null && paths.length > 0) {
     dragFiles = new File[paths.length];
     for (int i = 0; i < paths.length; i++) {
      StringpathName= tree.getPathName(paths[i]);
      dragFiles[i]=new File(pathName);
     }
     Transferable transferable = new FileListTransferable(dragFiles);
     dge.startDrag(null, transferable, this);
   } }}
```

This implementation is shown in Figure (5).

DropTarget  said to be "Drag Under" feedback to the user i.e the components are droppend into the GUI Interface.

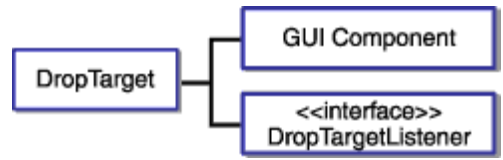

Figure 4. Drop Target Flow Review

**Drop methods:**

- **Validation**
- **Accepting the incoming drop**
- **Data transfer**

**Validation**

 First we have to do validation of DataFlavor before the drop operation. The DropTargetListener chooses the DataFlavor and checks the operation before the transfer takes place.If none of the DataFlavors and operations are acceptable, the DropTargetListener sends a rejectDrop message to the DropTargetDropEvent.The event object is depends on the drag source. Event object check with isLocalTransfer if the drag source is in the same JVM.depends on the event object the DropTargetListener can choose an appropriate DataFlavor.

 **Accepting the incoming drop**

 Once the drop passes the validation, the DropTargetListener sends an acceptDrop message  with the desired operation specified to the DragSourceListener in its dragDropEnd method. After the DropTargetListener accepts the drop, it retrieves the Transferable and requests its data.

**Data transfer**

Data transfer in the case of native-to-Java or inter-JVM transfers, the data's representation class should be a subclass of java.io.InputStream, since references to a Java object make sense only in the JVM.

Drag and Drop of reusable component can be it can basically be broken down into three main components.

1. Starting the drag where the drag action is recognized by the component
2. Converting the drag component into a transferable component
3. Dropping the transferable component into the drop target.

**Drag Methods**

The following Drag methods facilitates the event whether or not to accept reusable componets.

- **public void acceptDrag()**
- **public void rejectDrag()**

*public void acceptDrag(int dragOperation)*

This method indicates that the repository will accept the drag of reusable component.

*public void rejectDrag()*

This method indicates that repository will not accept the drag of reusable component.

**Drop Methods**

Similar to the drag events, we can accept or reject drops.

- **public void acceptDrop()**
- **public void rejectDrop()**

*public void acceptDrop(int dropAction)*

This method will accept a drop of type dropAction. If you decide to accept the drop, call this method, process the drop, and then call the dropComplete() method below.

*public void rejectDrop()*

This method rejects the drop of reusable component.

*public void dropComplete(boolean success)*

This method tells the source of the drag that the drop was completed. The success argument should be true if the drop of reusable component was successful, false otherwise.

```
// This method handles a drop for a list of files

    FileTree.FileTreeNode node =

(FileTree.FileTreeNode)treePath.getLastPathComponent();
    // Highlight the drop location while we perform the drop
    tree.setSelectionPath(treePath);
    // Get File objects for all files being
    // transferred, eliminating duplicates.
    File[] fileList = getFileList(files);
    // Don't overwrite files by default
    copyOverExistingFiles = false;
    // Copy or move each component object to the target
    for (int i = 0; i < fileList.length; i++) {
     File f = fileList[i];
     if (f.isDirectory())
     {
       transferDirectory(action, f, targetDirectory, node);
     }
    else {
     try {
     transferFile(action, fileList[i],
       targetDirectory, node);
       } catch (IllegalStateException e) {
        // Cancelled by user
        return false;
     }}} return true;

 }
```

**III. Results**

In this paper we describe a scenario where by the designer/user can build as much as possible of the custom application by interactively and graphically dragging components out of a smart object suitcase, specifying their exact behavior by directly editing their attributes ,then graphically connecting them to other objects by drag and drop, and reuse that dropped components[1].

Our GUI interface supports to drag the components from GUI interface to Logical cursor. The dropped component is reused with its interface specification and business logics.

Figure (5) represents dragging a java file from the Interface. and dropped into the local drive of the same system. our interface shows the hidden components also. So the user able to drop the hidden components also. This functionality said to be "Drag Over". The DragSource object manifests "Drag Over" feedback to the user, in the typical case by animating the GUI Cursor associated with the logical cursor.

The Drag Under functionality is shown in the Figure (6).This Figure shows a java file is dragged from local drive of the same system and dropped into our Interface. This interface is act as a repository to store the dropped components into that. The DropTarget object manifests

"Drag Under" feedback to the user, in the typical case by logical cursor associated with the animating the GUI Cursor.In the Figure (5) the system file is dragged from the Jal Drag Tree and Dropped into the Local Drive E.in the Figure (6) info file is dragged from the local drive D and dropped into Jal Drop Tree interface. Here we can edit the file names also and we can enable and disable the drag and drop operation using the check box provided in the interface.

A drop target accepts an incoming drag. The process of accepting the dragged information generates a series of events that we can respond to. Figure(6) represent the GUI with drop target.

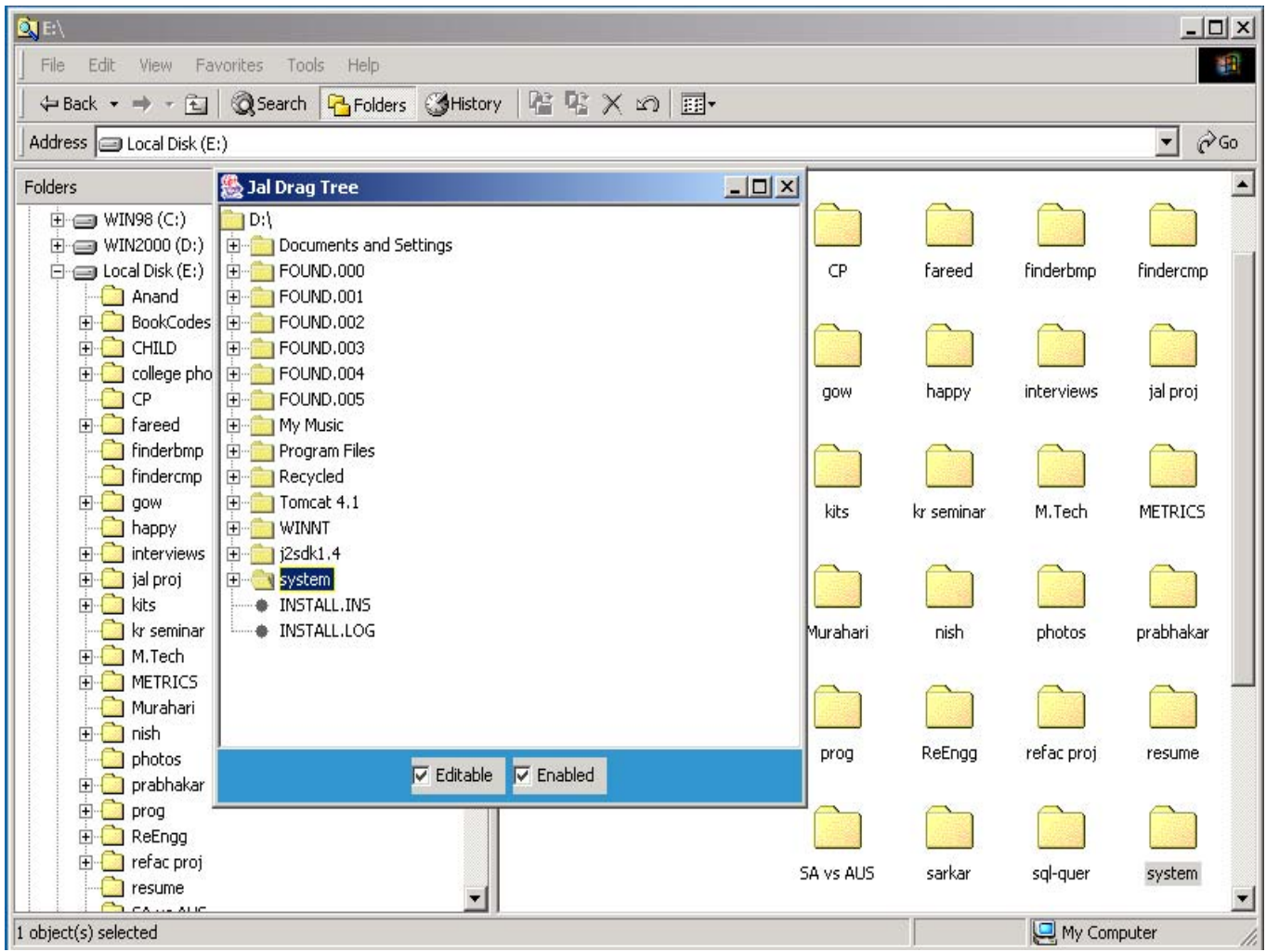

Figure 5. Screen shot of "Drag Over" mechanism.

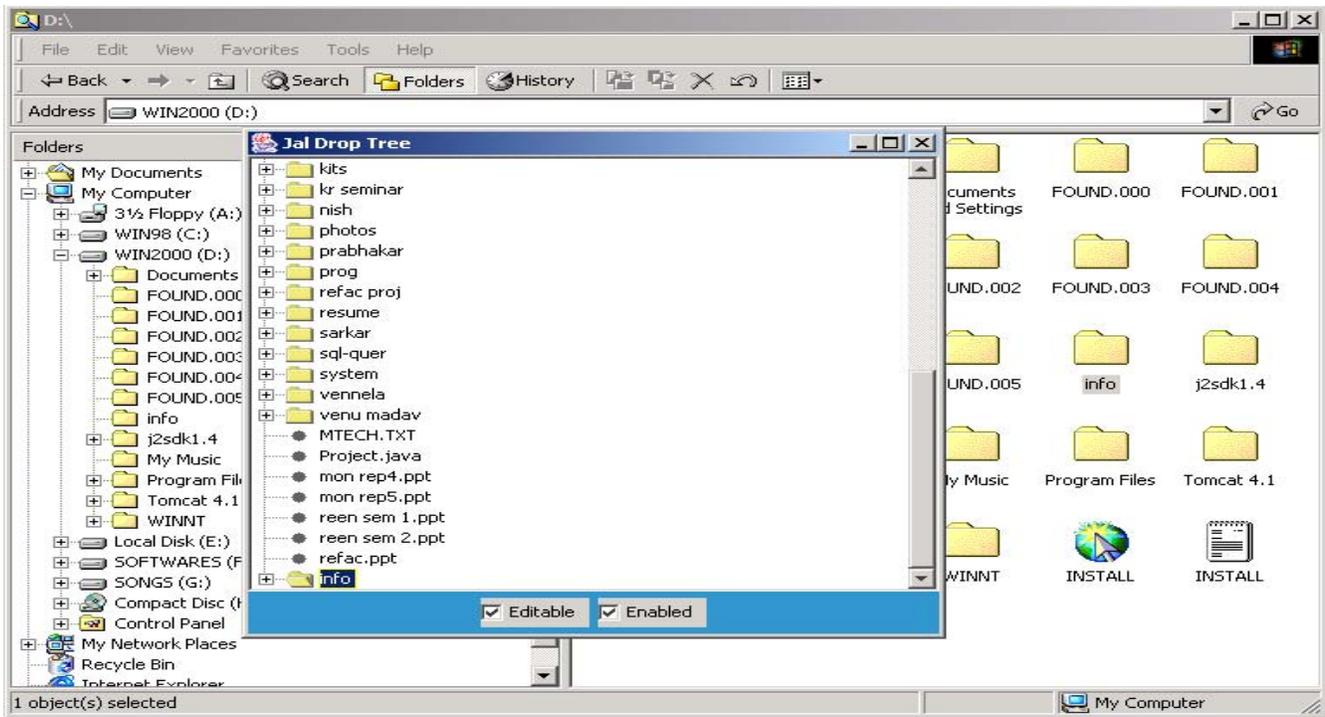

Figure 6. Screen shot of "Drag Under" mechanism.

## IV. Conclusion and Future Works

The ability to let users drag components from their Interfaces into our application is almost a requirement of a modern, commercial user interface [1]. Reusable components designed with "Drag and Drop" style have many advantages over designs based on the traditional "copying style," [4] including improved execution efficiency, higher reliability and enhanced reusability. Over the years drag-and-drop has gone from a cool feature to a required piece of most user interfaces.

Developing reusable components software requires a designer to determine how to structure a software system so as to achieve the necessary functionality, while at the same time increasing the reuse potential of the software. Component identification [6] centers on the application's domain, with reuse focusing specifically on an organization's future systems. The main approach is to categorize components, identify component boundaries, and specify where components are related.

Our repository is implemented to support process for component reuse using drag and drop mechanism. When the component developers perform component-based projects, our repository provides reuse facilities that are necessarily required during component-based development. Through the component repository with this mechanism components can be efficiently reused in component reuse process.

For future work we suggest to built up your own reusable component repository and provide the functionality to reuse the components using both drag and drop mechanism.

## AUTHORS PROFILE

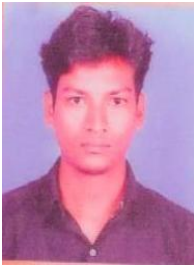

**B.Jalender** Received the Bachelor's Degree in Computer Science and Engineering from JNT University Hyderabad in 2003 and Master's Degree in Software Engineering from Kakatiya University Warangal in 2006.Now pursuing Ph.D in Computer Science and Engineering from Osmania University College of Engineering, Hyderabad. He is presently working as Assistant Professor in IT Department at VNR VJIET Hyderabad. His current research interests include software engineering especially in the areas of reusable software components and component based software engineering.

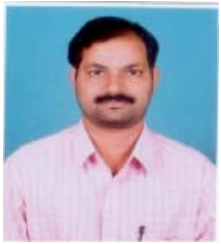

**Dr.A.Govardhan** did his BE in Computer Science and Engineering from Osmania University College of Engineering, Hyderabad, M.Tech from Jawaharlal Nehru University, Delhi and Ph.D from Jawaharlal Nehru Technological University, Hyderabad. He is presently working as Principal, JNTU Jagtial, Karimnagar, Andhra Pradesh. He has guided more than 100 M.Tech projects and number of MCA and B.Tech projects. He has 93 research publications at International/National Journals and Conferences. He has been a program committee member for various International and National conferences. He is also a reviewer of research papers of various conferences. He has delivered number of Keynote addresses and invited lectures. He is also a member in various professional bodies. His areas of interest include Databases, Data Warehousing & Mining, Information Retrieval, Computer Networks, Image Processing and Object Oriented Technologies.

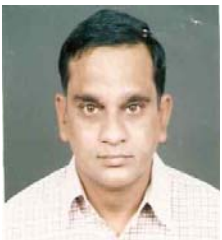

**Dr.P.Premchand** has graduated in Electrical Engineering from National Institute of Technology, Jamshedpur. He has obtained his M.E and Ph.D degrees in the branch of computer science and engineering from Andhra University, Visakapatnam.He has joined as lecturer in the department of CSE of Andhra University, Visakapatnam.Later he has shifted to Osmania University, Hyderabad into department of CSE as Associate professor. He has also served in various positions such as director at AICTE New Delhi and as an additional controller of Exams at Osmania University, Hyderabad. Later he is elevated as Professor in his parent Department at Osmania University and served as Head of the Department CSE and chair man Board of Studies and presently he is serving as Dean Faculty of Informatics at Osmania University Hyderabad. He is also an active member of AICTE –NBA, selection committee member at J.N.T.U, A.U, A.N.U, K.U, ISRO, NRSA and ADRIN. He is actively involved in research, supervising 5 research students for the award of their Ph.D and many more students are pursuing their Ph.D at O.U, JNTU and A.N.U. He has presented several papers in national and international conferences and journals.

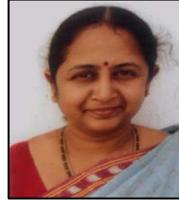

**Dr Kiran Mai, Cherukuri** did her graduate course in Electronics and Communication Engineering from Jawaharlal Nehru Technological University, Hyderabad and Masters in Software Systems from Birla Institute of Science and Technology, Pilani. From 1997 she is working in the Department of Computer Science, VNRVJIET Hyderabad. Currently she is the professor in the department. Before joining the teaching profession, she worked 7 years in industry where she was handling software projects in 'C', COBOL, Visual Basic and ORACLE. She is a life member of ISTE. Her research interests include Databases, Data mining, Networks and Human Computer Interaction. is a member of the IEEE and the IEEE Computer Society.

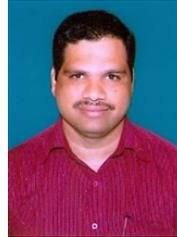

**G.Suresh Reddy** Received the Bachelor's Degree in Computer Science and Engineering from Bangalore University Bangalore in 1997 and Master's Degree in Information Technology from Punjab University Punjab. Now pursuing Ph.D in Computer Science and Engineering from National Institute of Technology Warangal. He is presently working as Associate Professor and HOD in IT Department at VNR VJIET Hyderabad. His current research interests include Semantic Grid, Data Warehousing and Mining, Text Mining and Information Security.